# Finite element 3-D model of a double quantum ring: effects of electric and laser fields on the interband transition

A. Radu[1*], C. Stan[1], D. Bejan[2]

[1] Department of Physics, Politehnica University of Bucharest, 313 Splaiul Independenței, Bucharest, RO-060042, Romania

[2] Faculty of Physics, University of Bucharest, 405 Atomiștilor Street, Măgurele, RO-077125, Romania

*Corresponding author: adrian.radu@physics.pub.ro

## Highlights

- 3D FEM model of a GaAs/AlGaAs droplet epitaxy DQR
- Effects of static electric and nonresonant laser fields
- Numerical 3-D laser dressing of the confinement potential
- Strongly anisotropic electron / hole pair polarizability
- Interband oscillator strength and recombination time

## Graphical abstract

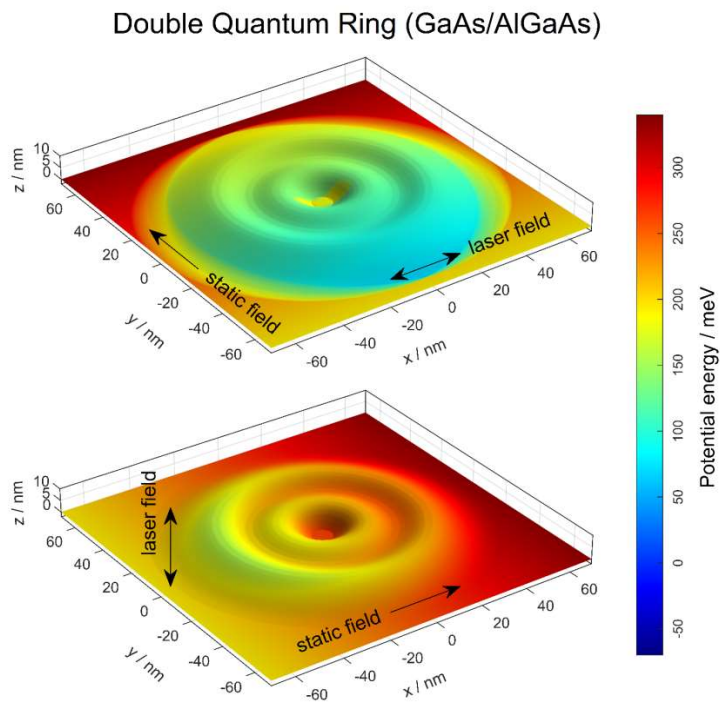

## Abstract

In this work, the changes in the energy of electrons and holes, oscillator strength and interband transition time when external fields are applied to a GaAs/AlGaAs semiconductor double ring grown by the droplet epitaxy technique are theoretically analyzed. We consider a static electric field and an intense laser field nonresonant with the quantum structure, with variable intensities and orientations with respect to the symmetry axis of the quantum ring. In the formalism of the effective mass approximation for electrons and holes, the energies and wavefunctions were numerically computed using the finite element method implemented with an accurate three-dimensional model of the real quantum ring. Laser dressing of the confining potential was performed using the exact integration formula at each point. Our results show major differences between the effects of the two types of applied fields, caused mainly by the static electric-field-induced strong polarizability of the confined electron-hole pair. In addition, the effects of both fields exhibit strong anisotropy in the electronic properties as a result of the particular flattened geometry of the quantum ring. Proper combinations of field strengths and orientations are helpful in designing accurate tools for the sensitive manipulation of interband radiative properties.
**Keywords:** quantum ring, droplet epitaxy, finite element method, electric field, laser dressing, interband transition, oscillator strength, recombination time.

## 1. Introduction

An important category of semiconductor nanostructures is quantum rings (QRs), which are remarkable systems, both dimensionally and functionally. On the one hand, they are similar to quantum dots, especially if they have a relatively small radius, and on the other hand, they can behave like quantum wires in the form of a loop, especially if they have a relatively large radius. It can be said that they have intermediate dimensionality between quantum dots and wires; thus, in a certain sense, they are meso-dimensional structures. Viefers *et al.* synthesized in a well-received article from 2004 the most important knowledge and studies from that time about QRs, with a particular emphasis on the problems of the characteristic spectrum and periodicity of the persistent current [1]. A relevant book by Editor V. Fomin, in its second edition in 2018, comprehensively presented the most relevant theoretical and experimental investigations related to QRs [2]. Manufacturing methods involving molecular beam epitaxy, Stranski-Krastanov growth, and droplet epitaxy are presented, along with concepts such as quantum interference and persistent current, Aharonov-Bohm and Berry phase effects, concentric QRs and coupled QR – quantum dot chains, spin interference effects in Rashba QRs, and effects of electromagnetic fields on the QRs. From a functional point of view, QRs are of interest because of the significant influence that external electromagnetic fields can have on their optoelectronic properties [3]. Especially in the case of relatively large radii, where specific quantum phenomena would already be less obvious in the case of an ordinary quantum dot, QRs can register important and interesting effects of external electric or magnetic fields [4,5]. The charge carriers in QRs can be strongly delocalized over a relatively large perimeter and interact significantly with the applied fields [6]. Special quantum phenomena can be investigated in such nanostructures, such as the Aharonov-Bohm effect [7,8]. From an experimental point of view, the studied shapes and sizes are limited by the specific technical possibilities and the related works are often focused more on the descriptive side and details of the involved processes [9]. From a theoretical point of view, the optoelectronic behavior of QRs can be investigated more thoroughly, and predictions can be made about how these structures will interact with the electromagnetic field under a variety of physical conditions and geometrical configurations [10]. Unfortunately, many theoretical works that address these topics tend to use simplified geometric models, which can provide at most rough qualitative evaluations of some parameters of practical interest. One reason can be related to the temptation of easier exploitation of some possible semi-analytical solutions of the physical equations involved. Other reasons are the difficulty of accurate geometrical modeling of real nanostructures and the complexity of fully numerically solving the intervening three-dimensional (3-D) problems with partial derivatives. For example, theoretical models often resort to the adiabatic decoupling of the 3-D Schrödinger problem into a polar 2-D problem and an axial 1-D problem, using the fact that a QR is, in many cases, a relatively flat, wide, and slightly tall structure [11]. Thus, a QR becomes a 2-D quantum confinement problem, losing information about the real 3-D profile of these structures, which can only be roughly reflected in the 1-D auxiliary axial problem. In other words, a QR is considered to have a relatively wide toroidal structure, with a constant rectangular section and slightly high, in the axial direction. Moreover, if a QR has perfect cylindrical symmetry, the 2-D problem in the transverse plane can be further reduced to a 1-D radial problem. Another simplifying approximation that has been attempted for relatively large-diameter QRs is to consider it as a curved quantum wire and separate the 2-D problem in the axial section of the structure. The remaining problem is that of one-dimensional QR [12]. The results of such approaches can be of fundamental theoretical or even qualitative interest in certain practical situations, but they have important limitations in other situations of interest, for two reasons. First, the technological methods by which the QRs are obtained never favor radial profiles with constant height of the QRs, or a simple geometric shape. Second, although structural axial symmetry is practically achievable, any interaction with fields that have a radial or transverse component alters the physical symmetry and makes it impossible to reduce the mathematical dimensionality of the equations that describe the problem. These difficulties and shortcomings are overcome if a more realistic, 3-D model is used, with a more accurate description of the structural geometry and the concrete possibility of introducing terms of asymmetric interaction with the external fields. A consequence of this more precise approach, which can be seen as a disadvantage, is the need to use fully numerical calculation methods without analytical or semi-analytical parts of the solutions. The management of the method requires greater computing resources and makes it mandatory to resort to particular cases for comparison and validation, or access to experimental data to confirm sufficient calculation accuracy.

Several very recent studies have dealt with various physical processes in double quantum rings (DQRs): the influence of the electric field (EF) on excitons [13], the effects of the magnetic field and ring dimension [14], magneto-optical absorption [15], and optical rectification [16]. In this work, we theoretically investigate a DQR model inspired by a real semiconductor structure manufactured by another collective [17] through the droplet epitaxy technique. The formation of this structure favors an interesting DQR configuration. Even if its geometry is relatively complex and more technical to approach by numerical models, it has the advantage of describing an existent quantum structure. The immediate implication is that, unlike the case of simplified structures addressed by many theoretical analyses (spheres, cylinders, cones, disks, etc.), the theoretical predictions and the numerical results obtained in this work are more likely to be tested experimentally. Recently, we have also published a theoretical study on a realistic triple quantum ring [18]. In the present article we propose to estimate the extent to which the strength and energy of the interband transitions can be modified by the interaction of a DQR with two types of fields: a static EF and an intense laser field (LF) non-resonant with transitions specific to the structure. The effects of the two fields are studied with the appropriate theoretical models, regularly used in the literature: the EF introduces a linear term in the Hamiltonian and produces the static confined Stark effect; the LF is taken into account according to the laser dressing theory of the confinement potential. The latter is a particular case of the dynamic confined Stark effect which is obtained in cases of time-varying fields. The laser-dressing model is valid for nonresonant fields, at high laser intensities, and frequencies in the far-infrared range. The reason why the effects of these two types of fields are commonly studied for nanostructures are the quasi-static shifts of the energy levels. In this work, both interactions were studied at various orientations of the fields with respect to the axis of symmetry of the DQR. The difference in effects between the application of the axial and transverse fields was particularly observed. The proposed model and its predictions consider a sufficiently low-temperature regime for the population of the excited conduction and valence subbands to be practically negligible. In particular, with regard to the effect of the intense LF on the DQR, to the best of our knowledge, this is the first 3-D model that studies the exact laser dressing along a certain direction of a realistic quantum structure. We can expect that some qualitative aspects highlighted in this study can be extrapolated to other rings, with different profiles and materials. We can mention here the tunability of the interband emission with the use of an axially polarized nonresonant LF and the effective control of the recombination time by using an EF oriented in the plane of the rings.

The remainder of this paper is structured as follows: Section 2 presents the theoretical concepts and the calculation model used; Section 3 contains the results obtained and their analysis; and some conclusions are formulated in Section 4.

## 2. Theory

The quantum system we theoretically investigated is the GaAs/AlGaAs semiconductor DQR produced by droplet epitaxy, as presented in detail in Refs. [17-21]. Figure 1 illustrates the geometry of this axially symmetrical nanostructure, which was recreated based on the original article by Kuroda *et al.* [17]. The contour of the axial section presented in Fig. 1a is formed by a radially oriented straight-line segment (blue line) and a profile curve $g(\rho)$ (red line). By rotating the contour around the axis of symmetry $z$, the surface of the DQR is obtained, that is, the interface between GaAs and AlGaAs materials.

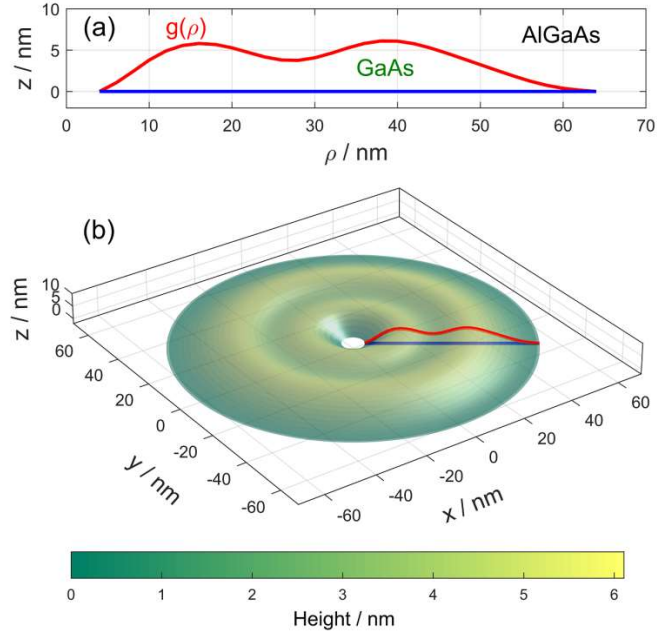

**Figure 1.** a) Axial section contour of the DQR; b) Interface between the two semiconductor materials viewed as the surface of revolution generated by revolving the axial contour about the $z$ axis.

We aim to study the precise way in which applied external fields change the ground energy levels of electrons and holes confined in this structure as well as the consequences of these changes on the interband transition. Owing to the relatively broad structure, the effective mass approximation of the two semiconductor materials is a reasonable model. Because the DQR has axial symmetry, the effective masses and confining potentials of the electron (e) and hole (h), respectively, are functions of the radial $\rho = \sqrt{x^2 + y^2}$ and axial $z$ coordinates:

$$m^*_{e,h}(\rho,z) = m^{*\,GaAs}_{e,h} + \left(m^{*\,AlGaAs}_{e,h} - m^{*\,GaAs}_{e,h}\right)\left[H(-z) + H(z - g(\rho))\right], \qquad (1)$$

$$V_{e,h}(\rho,z) = U_{e,h}\left[H(-z) + H(z - g(\rho))\right], \qquad (2)$$

where $H$ is the Heaviside step function, and $U_{e,h}$ are the conduction and valence band discontinuities at the interface between the two semiconductors, respectively.

The DQR-confined single-particle states of electrons and holes can be spatially described by state functions that verify the Schrödinger formalism. Electrically charged particles will be considered under the simultaneous actions of a constant EF $\mathbf{F} = \mathbf{u}_f F$ and a non-resonant LF of amplitude $\mathbf{A} = \mathbf{u}_p A$ of the vector magnetic potential, where the unit vectors $\mathbf{u}_f$ and $\mathbf{u}_p$ correspond to the EF direction and the LF polarization, respectively. If the values of the applied static EF are not too high, the quantum Stark effect can be studied in the approximation of quasi-bound states, with the energy eigenvalues and related state functions having quasi-zero imaginary parts [22,23]. It is also known that for sufficiently large values of angular frequency $\omega$ of the laser wave, the quantum states of the confined particles are quasi-stationary and can be calculated using a modified atemporal form of the Schrödinger equation [24]:

$$-\frac{\hbar^2}{2}\nabla\left[\frac{1}{m^*_{e,h}(r)}\nabla\psi_{e,h}(r)\right] + \tilde{V}_{e,h}(r)\psi_{e,h}(r) - q_{e,h}r\cdot u_f F\psi_{e,h}(r) = E_{e,h}\psi_{e,h}(r), \qquad (3)$$

where

$$\tilde{V}_{e,h}(r) = \frac{\omega}{2\pi}\int_0^{\frac{2\pi}{\omega}} V_{e,h}\left(r + \boldsymbol{\alpha}_{e,h}(r)\sin(\omega t)\right)dt \qquad (4)$$

is the laser-dressed confinement potential, $q_{e,h} = \pm e$ are the electric charges of the particles, and $\boldsymbol{\alpha}_{e,h}(r) = \frac{q_{e,h}\boldsymbol{A}}{\omega m^*_{e,h}(r)}$ are the corresponding vector laser parameters. It is observed that the scalar laser parameters $\alpha_{e,h} = |\boldsymbol{\alpha}_{e,h}|$ are not independent, because $\alpha_h = \alpha_e \frac{m^*_e}{m^*_h}$ at any point. A "high enough" frequency is the approximation made in the high-frequency Floquet theory (HFFT), which is the basis of the theoretical model of laser dressing [25]. The original approach was intended for atomic systems and later extended to hydrogenic impurities in semiconductors and any confinement potential, including quantum wells, wires, and dots. The zero-order approximation of the HFFT is generally valid for a non-resonant laser field if the high-frequency condition $\hbar\omega \gg W$ is satisfied, where $W$ is the typical excitation energy from a given state [25,26]. In addition, in semiconductor nanostructures, an upper frequency limit must be imposed, such as the resonant interband transitions are not stimulated by the laser. In GaAs nanostructures and for laser frequencies of tens of THz the photon energy generally satisfy the high-frequency condition and is also below the bandgap of the well material [26].

The fields with general orientations $u_f$ and $u_p$ can be expressed in terms of Cartesian components. Given the axial symmetry of the DQR, we can arbitrarily choose the x-axis as the direction of the transverse component of the laser parameter. Thus, we have $\mathbf{F} = F\sin\theta\cos\varphi u_x + F\sin\theta\sin\varphi u_y + F\cos\theta u_z$ and $\boldsymbol{\alpha}_{e,h} = \alpha_{e,h}\sin\gamma u_x + \alpha_{e,h}\cos\gamma u_z$, where $\theta$ and $\gamma$ are the polar angles of the EF and LF, respectively, and $\varphi$ is the azimuthal angle of the EF.

The dressed potential given by Eq. (4) was evaluated by numerical integration at each point of interest, and Eq. (3) can be solved using a 3-D numerical finite element method (FEM). The dependence on the parameters $F$, $\alpha_e$, $\theta$, $\gamma$, and $\varphi$ of the normalized state functions $\psi_{1,e,h}$ and of the eigenenergies $E_{1,e,h}$ of the ground states of the particles was thus determined.

The linear term in $F$ of the Hamiltonian in Eq. (3) makes the Stark problem amenable to axial/transverse decoupling for sufficiently small ground energy levels relative to the confining potential depth. Therefore, it is expected that the Stark shift at some orientation of the EF is the sum of the energy variations obtained in the separate axial/transverse problems. If this prediction is confirmed by the results, it may be sufficient to study the two separate Stark problems in conjunction with various laser polarization orientations. The problem of the laser dressing of the confining potential is completely different. The dependence of the Hamiltonian on the laser parameter is nonlinear, therefore the axial/transversal decoupling of the LF problem is not predictable. A separate study on the variation in the solutions with the orientation of the polarization direction may be necessary.

The intensity of the interband optical absorption/emission line is given by the oscillator strength [27]:

$$f = \frac{E_P}{2E_t}\left|\iiint \psi_{1,e}(r)\psi_{1,h}(r)d^3r\right|^2, \qquad (5)$$

where $E_P$ is the Kane energy, $E_t = E_g + E_{1,e} + E_{1,h}$ is the transition energy, and $E_g$ is the energy gap of the DQR inner material.

The interband radiative lifetime may also be relevant for experimental studies [28,29]:

$$\tau = \frac{6\pi\varepsilon_0 m_0 c^3 \hbar^2}{e^2 n f\, E_t^2}, \qquad (6)$$

where $\varepsilon_0$ is the vacuum permittivity, $m_0$ is the free electron mass, $c$ is the speed of light in vacuum, and $n$ is the refractive index of the semiconductor.

## 3. Results and discussion

For the numerical calculations we used the following values of the material parameters: $m_e^{*GaAs} = 0.067m_0$, $m_h^{*GaAs} = 0.51m_0$, $m_e^{*AlGaAs} = 0.093m_0$, $m_h^{*AlGaAs} = 0.57m_0$ [30], $U_e = 262$ meV, $U_h = 195$ meV [31], $E_p = 22.7$ eV [32], $E_g = 1.51$ eV, $n = 3.26$ [33].

Under the given field conditions, the potential energy of charge carriers, that is, $\tilde{V}_{e,h}(r) - q_{e,h} r \cdot u_f F$, is a function of three scalar variables (Cartesian coordinates) that is difficult to fully visualize graphically. However, the potential energy can be graphically represented by a color gradient on certain surfaces included in the spatial domain of the model. Figure 2 shows chromatic representations of the potential energy of the confined electron at the points of the upper interface $z = g(\sqrt{x^2 + y^2})$ of the quantum structure (Fig. 2a,c,e,g), as well as in a transverse section plane (Fig. 2b,d,f,h). According to Fig. 1a, the section plane $z = 5$ nm can highlight both rings of the quantum structure. For the hole, the graphs are qualitatively similar but with a different value of the potential barrier and much smaller potential changes in the LF, owing to the very different laser parameter values. Figures 2a,b and 2c,d correspond to a transverse laser dressing on the axis of symmetry of the DQR in the absence of the static EF, with $\alpha_e = 10$ nm and $\alpha_e = 20$ nm, respectively. It is observed that although the perimeter of effective confinement expands along the $x$ direction of the laser polarization, the depth of effective confinement generally decreases, leaving only four deeper zones aligned along the $y$ direction in the vicinity of the rings. As the laser parameter increased, these deep confinement zones narrowed. Figures 2e,f refer to the simultaneous presence of the two types of fields, in the $x$ direction.

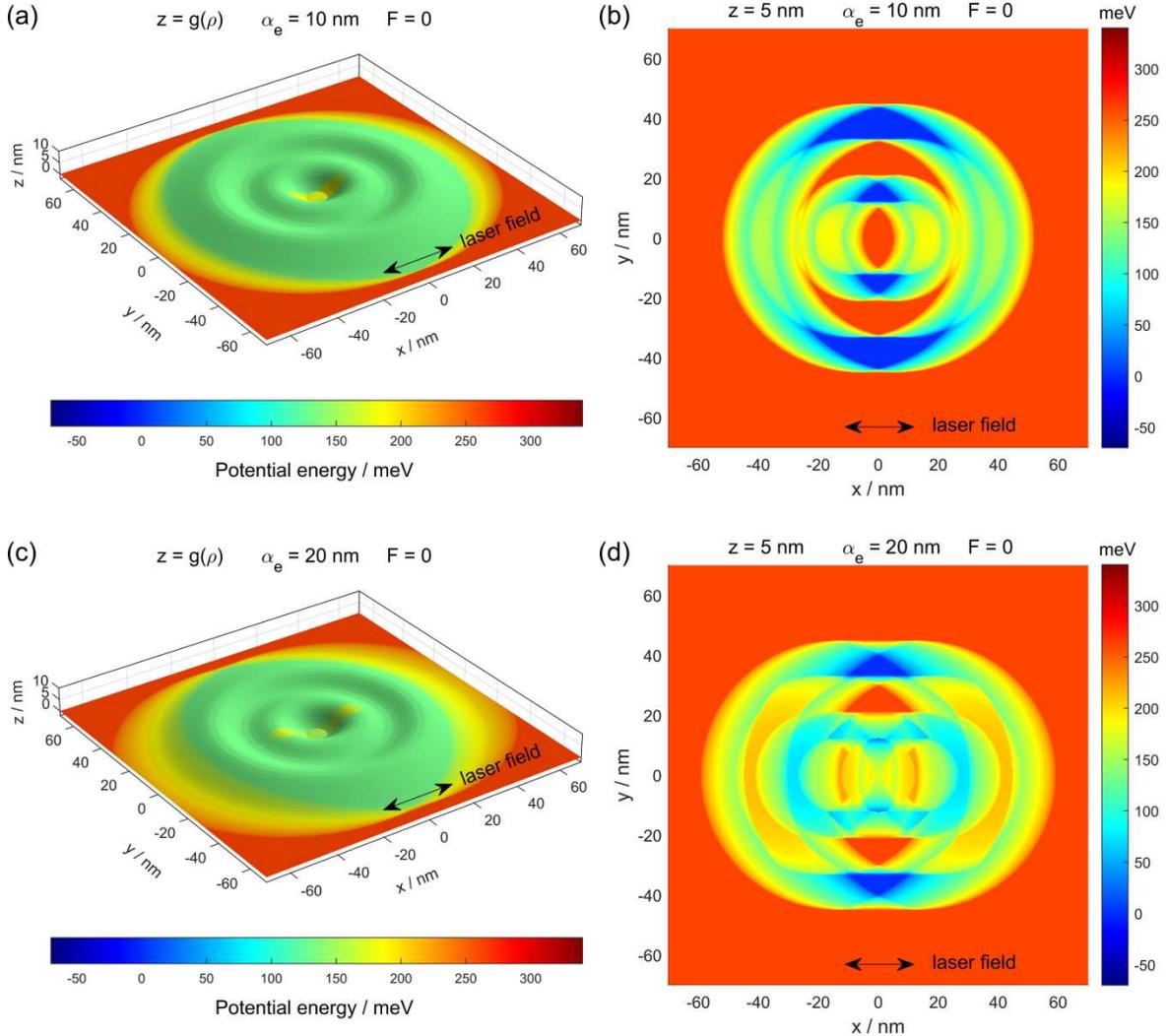

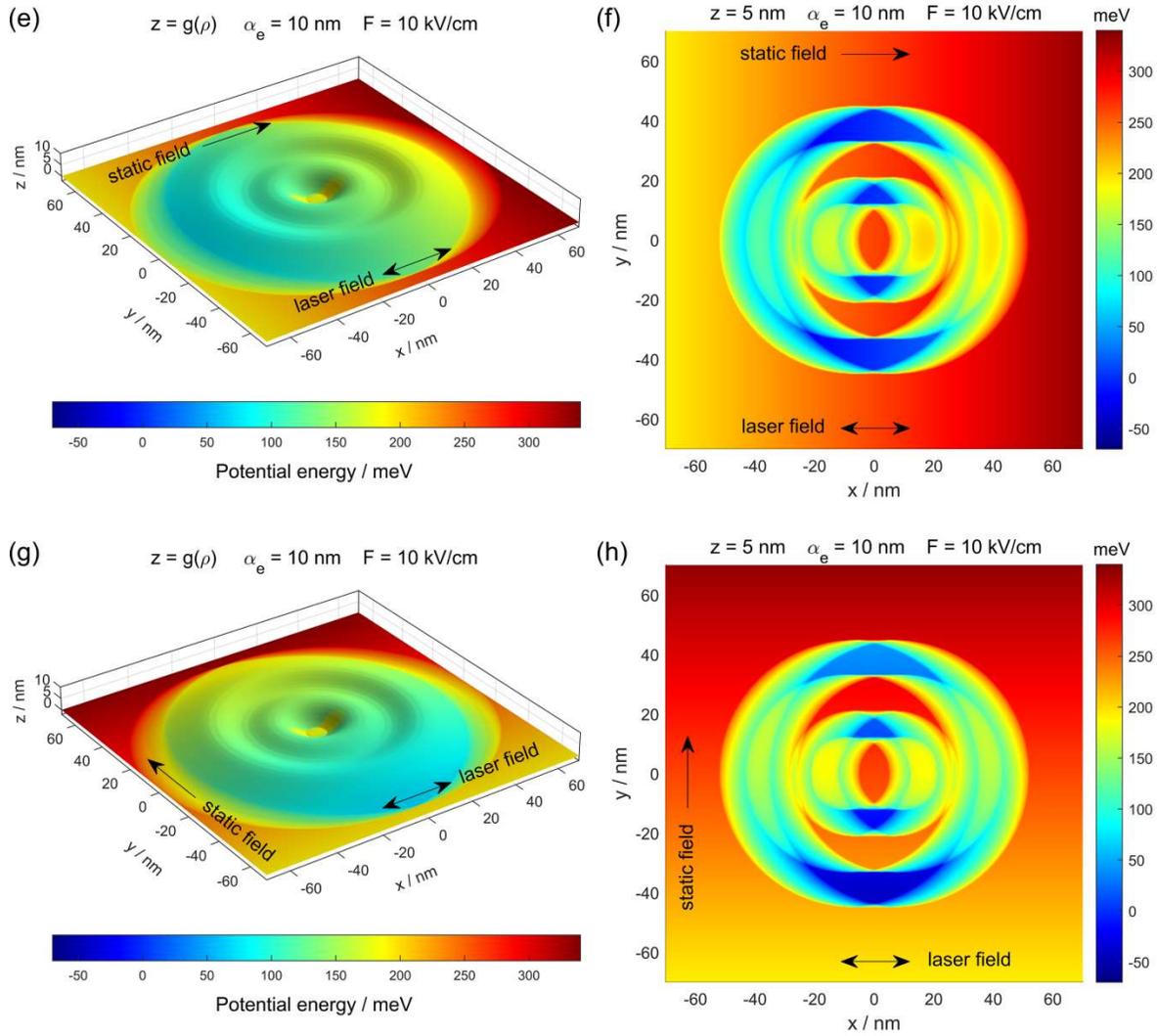

**Figure 2.** Electron potential energy on the upper surface of the DQR (a,c,e,g) and in the transverse section plane $z = 5$ nm (b,d,f,h), for some particular cases of transverse laser and EFs.

Comparing Fig. 2b and Fig. 2f, which have the same value of the laser parameter, the "inclination" in the $x$ direction of the entire potential profile is noticeable, which is a behavior characteristic of the action of the static EF. Figures 2g,h correspond to the case with laser dressing in the $x$ direction and EF in the $y$ direction.

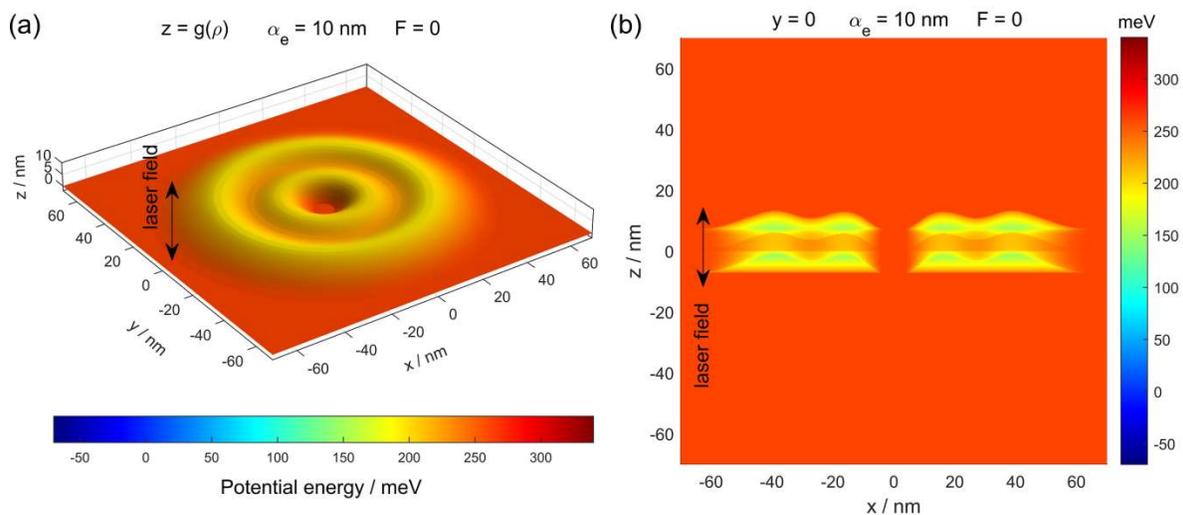

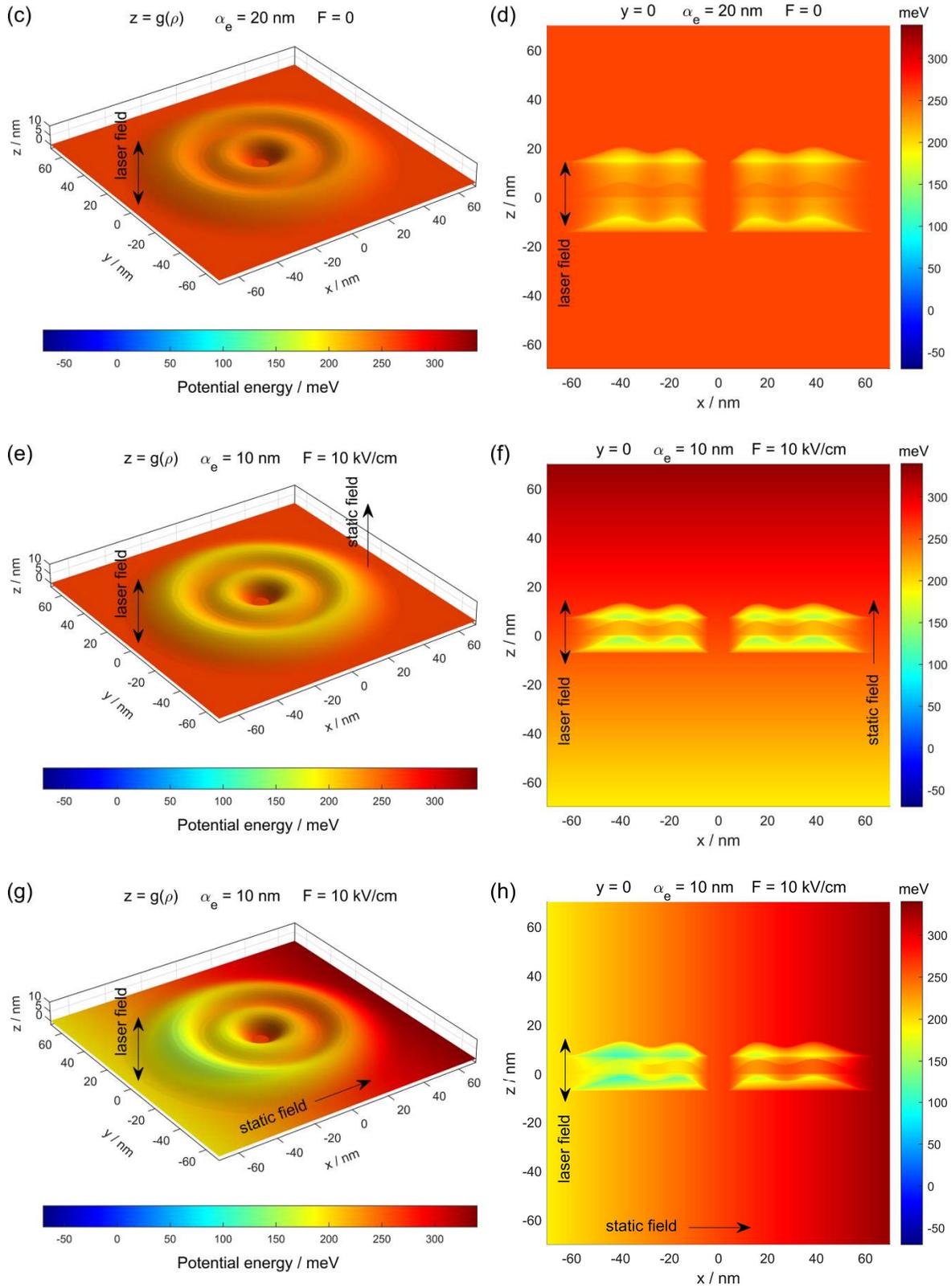

**Figure 3.** Electron potential energy on the upper surface of the DQR (a,c,e,g) and in an axial section plane (b,d,f,h), for some particular cases of laser and axial EFs (except for g,h, where the EF is transverse).

Figure 3 shows graphs similar to those in Fig. 2, but with an axial laser dressing. Except for the case in Fig. 3g,h, where the static field is applied perpendicular to the axis of symmetry, all situations presented here preserve the axial symmetry of the confinement. The appearance of a double potential well in the $z$ direction is a characteristic known from the behavior of quantum wells when dressed with laser parameters that exceed the typical width of the confinement zone [34]. A similar behavior of the laser-dressed confinement potential was also reported for quantum wires [35-37] and quantum dots [38].

The energies and wavefunctions (WFs) of the particles in their ground states were calculated under different field conditions using the FEM applied in a 3-D domain including the DQR. The geometry of the computational model used a Cartesian coordinate system, whose center was considered at the center of the plane base of the DQR. The spatial domain of the model was cylindrical in shape, coaxial with the DQR, having a radius of 130 nm and a height of 96 nm and extending along the axial $z$ direction between coordinates $-46$ nm and 50 nm. Thus, the GaAs inner region was approximately in the center of the working domain. A Neumann-type condition was used for the boundary of the cylindrical domain, canceling the flux of the WF. We worked with parametric solutions by varying the values of the applied fields and their orientations with respect to the axis of symmetry, $z$. The mesh of the FEM was of the free tetrahedral type, with refinement adapted to the DQR geometry. More technical details of the mesh are listed in Table I. The mesh used was more refined inside and near the GaAs volume, where the probability of particle localization was relatively high. Therefore, approximately 39% of the vertices belong to the spatial domain of the DQR, whose volume represents only approximately 0.85% of the total volume of the model domain. The element volume ratio in Table I represents the ratio of the largest and smallest volume elements in a given domain.

**Table I.** Tehnical details of the FEM 3D mesh

| Spatial domain | Mesh vertices | Tetrahedra | Element volume ratio ($10^{-9}$) | Mesh volume (nm$^3$) |
|---|---|---|---|---|
| Entire volume | 169658 | 976127 | 8.611 | 5094000 |
| DQR volume | 65630 | 246568 | 86.25 | 43410 |

The volumetric 3-D representation of the entire mesh is not practically feasible; however, one can visualize the mesh on the cylindrical boundary of the numerical model and on the interface between GaAs and AlGaAs, as shown in Fig. 4. The color scale encodes the size of the surface finite elements.

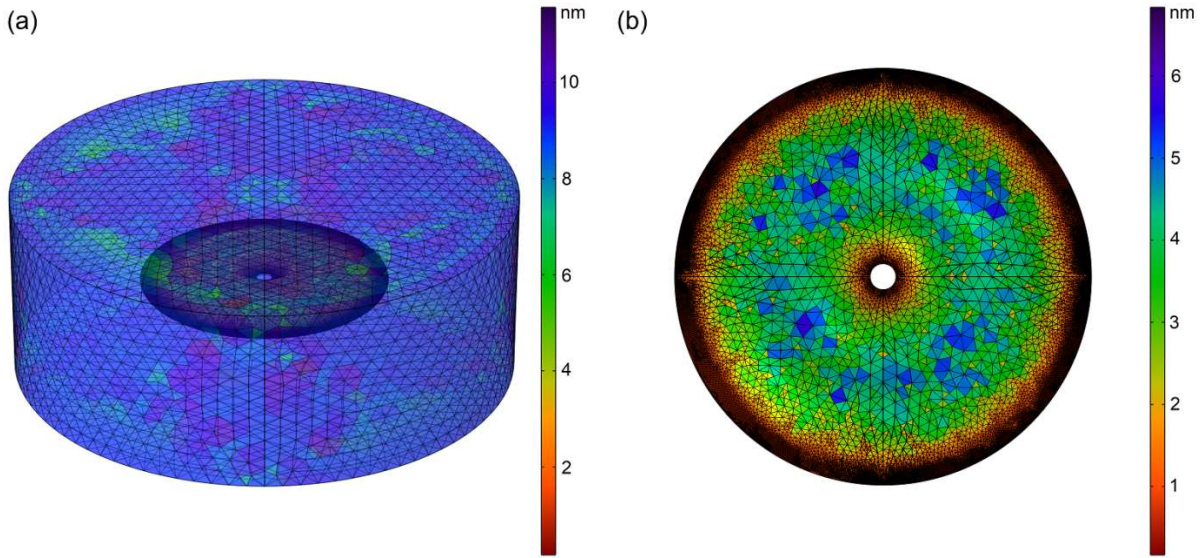

**Figure 4.** (a) Overview of the cylindrical boundary mesh; (b) Details of the mesh on the upper surface of the DQR.

Figure 5 shows how the electron and hole WFs change in the presence of each of the two types of applied fields. In the presence of the transversely polarized LF in the $x$ direction (Fig. 5a), both the electron and the hole form maxima of the WFs symmetric with respect to the $x$ axis and aligned along the $y$ axis, in agreement with the observation of deep confinement zones in Figs. 2b,d. For the electron, a slight tendency to shift the maxima from the outer to the inner ring is observed at $\alpha_e = 20$ nm. It should be noted that this double confinement behavior is more obvious for the electron because, as shown above, the laser parameter of the hole is much lower than that of the electron in the same laser wave. In the static EF oriented along the positive $x$ direction (Fig. 5b), a displacement of the maxima of the WFs along the field, in opposite directions, is obtained. The spatial separation of the two particles was fast, occurring for relatively low values of the applied EF. The large diameter of the DQR favors this extensive migration of the two particles towards diametrically opposite points of the outer ring, generating a field-induced strong polarizability. Figure 5c shows how the WF of the electron changes in an axially polarized LF, with the formation of the characteristic maxima of the double confinement along the polarization direction, at values of the laser parameter higher than the typical axial size of the DQR. However, the WF retains its full axial symmetry. In the case of the hole, the changes in the axial LF of the WF were too small to be illustrated here. In addition, in the axial EF, the shifts along the $z$-axis of the maxima of the electron and hole WFs are small in absolute value because of the much stronger confinement in the axial direction and are not plotted here. However, the energies of confined particles can vary considerably with EF. It is worth noting that in EF the WFs of the particles do not significantly penetrate the semiconductor barrier material, even if they can be displaced considerably inside the structure in certain directions. In contrast, in a non-resonant LF, the WFs can extend far into the barrier material.

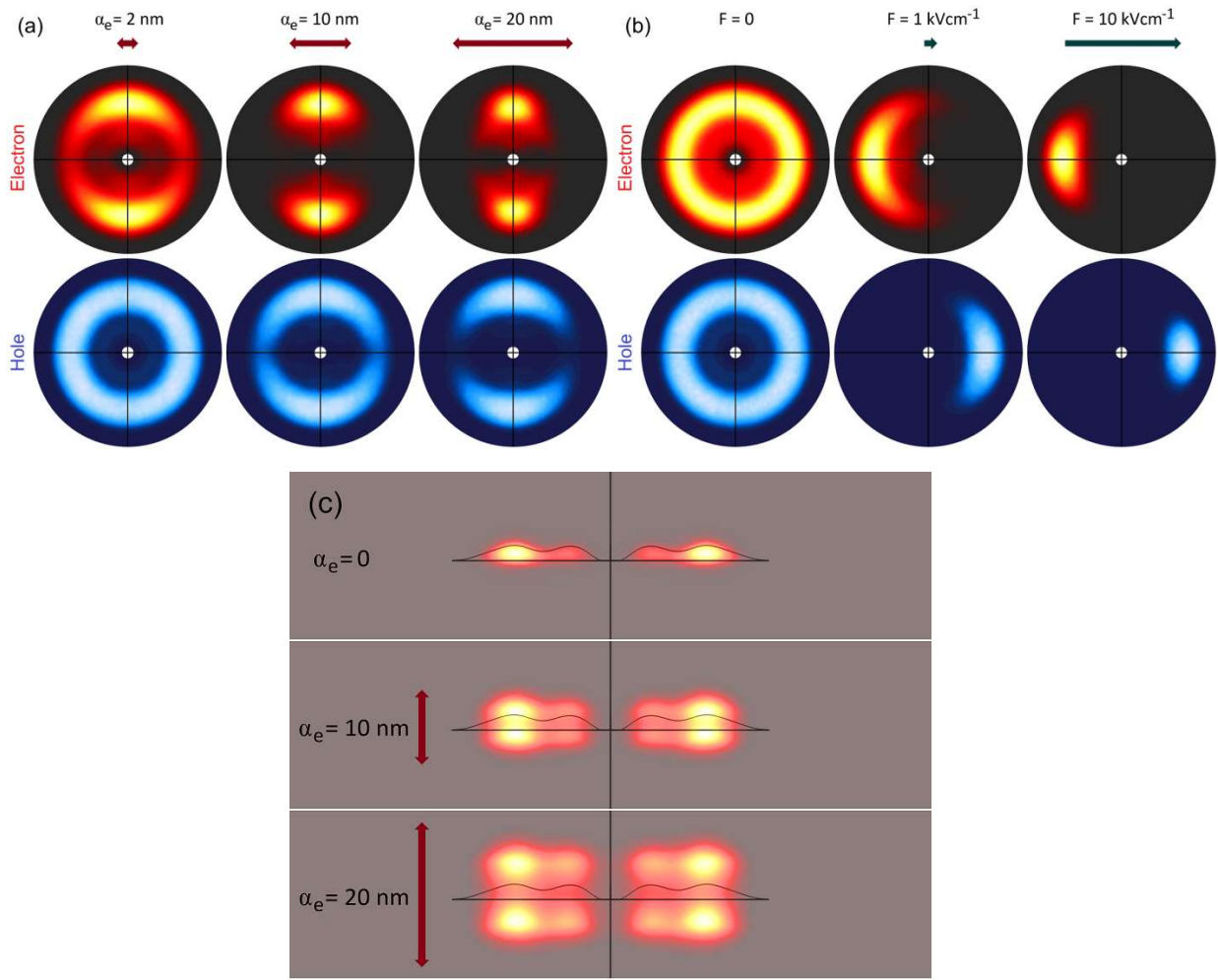

**Figure 5.** a) Unnormalized WFs of the electron and the hole in the cross section, laser dressed transversely in the $x$ direction, at three values of the laser parameter; b) The same type of representation, but with a static EF in the $x$ direction; c) Axial laser dressing of the electron's unnormalized WF, seen in axial section.

The obtained results show that the quantum nanostructure behaves very differently depending on the orientation of the applied fields, which is a direct consequence of the strongly flattened geometric shape. A quantitative measure of how electrons and holes change their localization inside such quantum structures when external fields are applied is shown in Fig. 6. The behavior of the particles is shown in a nonresonant LF (Fig. 6a) and static EF (Fig. 6b). The two main orientations, axial and transverse, of the static EF and the polarization direction of the laser wave were comparatively considered. On the same vertical scale, the subunit numerical values represent the localization probabilities in the GaAs volume of the electron and hole in the ground state, as well as the overlap integral of the normalized WFs of the two particles. Referring to Fig. 6a, we observe that in the case of transverse laser polarization, the localization probability inside the DQR of the electron decreases very little with the laser parameter, from approximately 88% to 87%, while that of the hole remains practically constant and equal to approximately 98%. The explanation for the difference observed in the absence of the field is related to the stronger confinement of particles with a higher effective mass. The reason for the different evolutions in the LF is also related to the effective mass, which determines the much smaller laser parameter of the hole. This aspect is much more obvious for the axial orientation of the laser polarization, in which case the laser dressing of the nanostructure potential has a strong effect on the electron WF. This different behavior in the axial case compared with the transverse case is explained by the typical size of the DQR in the two main directions. In the radial direction, the structure is very wide with respect to the values of the electronic laser parameter taken into account ($\alpha_e \leq 20$ nm), whereas in the axial direction, the average thickness of the DQR is of the same order of magnitude or smaller than the laser parameter. In the case of the hole, the difference between the axial and transversal behaviors is less obvious because the laser parameter values of the hole ($\alpha_h < 2.7$ nm) are small, even compared to the typical axial dimension of the DQR. In Fig. 6a we present for comparison the values of the laser parameters of both particles. Because they are directly proportional, for the unity of presentation in the subsequent figures, we use only the electronic laser parameter values as a reference. It is observed that the electron localization probability in GaAs decreases significantly and nonlinearly with the axially applied LF from 88% to approximately 5% near the maximum value $\alpha_e = 20$ nm. At first sight, this behavior seems counterintuitive because it translates into predominant localization in the barrier material. However, it is a well-known effect from the study of the laser dressing of carriers in 1-D or 2-D simple quantum well structures. Strong laser dressing considerably changes the spatial dependence of the effective confinement potential and, in the case of laser parameters exceeding the half-width of the quantum well in the laser polarization direction, leads to the appearance of a double-well-like confinement [34-37]. A hill-like potential barrier appears at the actual position of the well material, which explains the reduced probability of particle confinement in this area. These aspects can be intuitively understood from Figs. 2b,d,f,h. The process does not occur similarly for the transverse polarization direction, where at the same LF, the dressing is moderate relative to the diameter of the DQR. Regarding the spatial overlap of the WFs of the two particles, also in Fig. 6a, we observe an interesting aspect. The overlap integral started at approximately 94% in the absence of a field. At small values of the laser parameter ($\alpha_e \leq 5$ nm, i.e., below the typical axial size of the structure), the overlap is somewhat better in the case of axial polarization, because a moderate axial dressing tends to effectively colocalize the particles in the ground states [39]. This explanation is related to another effect known from the laser dressing of 1-D structures, namely, the reduction of the effective confinement width of low-energy states in quantum wells. However, at high values of the laser parameter ($\alpha_e > 5$ nm), the overlap of the electron and the hole strongly decreases with the axially polarized laser because of the rapid delocalization of the electron from the region of the inner material of the structure, whereas the hole changes its spatial extent only slightly. At the maximum value of the laser parameter considered, the overlap integral reached approximately 15%. The mixed-monotonicity aspect of the overlap variation for transverse polarization is less intuitive, showing a local minimum near $\alpha_e = 4$ nm and a local maximum at $\alpha_e = 12$ nm, which are characteristics of the DQR radial geometry. However, it was observed that this variation was less extensive, and the overlap remained almost permanently greater than 80%. Figure 6b shows the variations obtained with static EF. This figure uses two vertical axes because one of the curves shown (the overlap integral for the transverse EF) has a much larger variation than the other five. We must not lose sight of the fact that the probability of locating the particles inside the DQR is a global quantity, which is the result of an integral calculation. The EF, although it greatly influences the spatial distribution of the WFs inside the DQR, as seen in Fig. 5, changes only slightly the total localization probability inside the inner material of the nanostructure, for both the electron and the hole, at both main orientations of the field. In other words, the particles are forced by the field to move into the confining volume, but do not leave the DQR. This could still occur at much higher values of the applied static field, which is a separate tunnel effect problem [40,41]. With respect to the overlap integral, we have very different behaviors axially and transversely. In the case of the axial EF, the effect is weak because the structure is thin and the EF, regardless of its sign, fails to spatially separate the electron from the hole. The WFs are slightly compressed, in opposite directions, towards the potential barrier constituted by the AlGaAs material but remain spatially colocalized in the DQR, with an overlap integral of approximately 94%, very weakly decreasing with increasing field. The behavior of the particles in the transverse EF is completely different. Regardless of the direction in which it is applied, EF reduces the particle overlap from 94% to practically zero within a range of 1 kVcm$^{-1}$. The reason can be easily inferred from Fig. 5b: the electron and the hole are forced to move in the direction of the EF to diametrically opposite points of the DQR; the effect is very strong because the nanostructure has a large diameter. An important partial conclusion that emerges from the analysis of Fig. 6 is the highly anisotropic behavior of charge carriers, both in the non-resonant LF and static EF. The anisotropy of the electronic properties is directly determined by the strongly flattened geometric configuration of DQR. The radiative properties of the structure are expected to respond to external fields in a highly anisotropic manner.

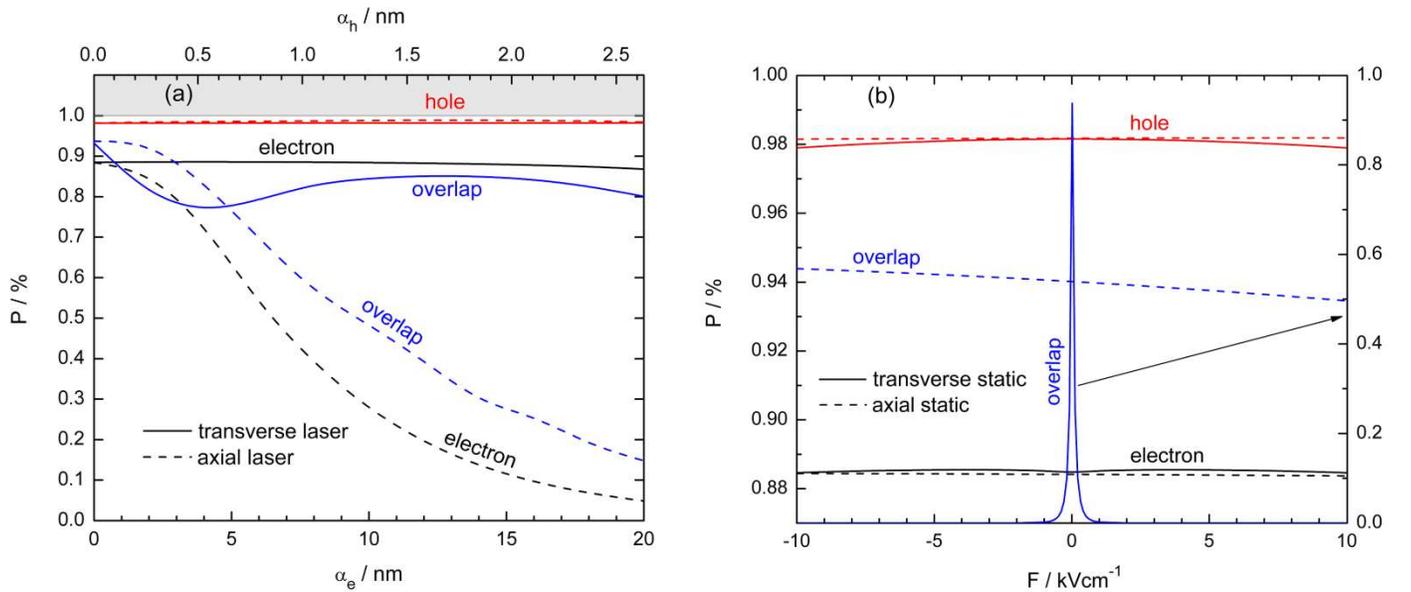

**Figure 6.** The probability of localization inside the DQR of the electron and the hole in the ground state and the overlapping integral of the two particles, as functions of the laser parameter (a) and the strength of the static EF (b). For both fields, the main axial and transverse orientations were considered.

The energy levels of the confined electron and hole ground states, measured relative to the GaAs conduction-band minimum and valence-band maximum, respectively, vary with the applied fields, leading to a change in the interband transition energy. Therefore, it is of interest to determine how their variation occurs when the intensity and orientation of the laser and static EFs are changed. It is relevant to first study the extent to which the effects of fields applied in some directions are deducible from the separate effects of the field components applied in the main (axial and transverse) directions. Tables II and III present the energy shift values of the electron and the hole at several values of the laser parameter and of the static EF, respectively, which are oriented at 45° to the axial direction of the DQR. The listed values of the fields in the tilted orientation case were derived from the vector sum of the axial and transverse field components. Table II shows in the red/blue columns that the shift of the electron/hole energy levels when the laser polarization direction is tilted with a polar angle of 45° is not simply the sum of the energy shifts obtained when dressing the DQR with the laser parameter components along the axial and transverse directions. This numerical result confirms what was theoretically anticipated in the preceding paragraph of Eq. (5). In the case of EF, the situation is the opposite, as the results in Table III demonstrate, within the limits of some small numerical differences. The linearity of the dependence of the Hamiltonian on the EF strength allows decoupling of the problem of a tilted EF into two separate problems, corresponding to the two main directions, which is not the case for the dressed potential.

**Table II.** Energy shifts of the electron and the hole, at several values of the electron laser parameter and three relevant orientations of the laser polarization direction

| | Laser dressing energy shift of the ground state / meV | | | | | | | | |
|---|---|---|---|---|---|---|---|---|---|
| $\alpha_e$ (nm) | $\gamma = 45°$ (tilted) | | $\alpha_e$ (nm) | $\gamma = 0°$ (axial) | | $\gamma = 90°$ (transverse) | | axial + transverse | |
| | electron | hole | | electron | hole | electron | hole | electron | hole |
| $2\sqrt{2}$ | 31.48 | 0.79 | 2 | 31.35 | 0.78 | 1.14 | 0.00 | 32.49 | 0.78 |
| $10\sqrt{2}$ | 126.20 | 7.94 | 10 | 125.57 | 7.89 | 4.26 | 0.15 | 129.83 | 8.04 |
| $20\sqrt{2}$ | 149.32 | 35.37 | 20 | 148.89 | 35.00 | 8.99 | 0.32 | 157.88 | 35.32 |

**Table III.** Energy shifts of the electron and the hole, at several values and three relevant orientations of the EF

| | Stark energy shift of the ground state / meV | | | | | | | | |
|---|---|---|---|---|---|---|---|---|---|
| $F$ (kV/cm) | $\theta = 45°$ (tilted) | | $F$ (kV/cm) | $\theta = 0°$ (axial) | | $\theta = 90°$ (transverse) | | axial + transverse | |
| | electron | hole | | electron | hole | electron | hole | electron | hole |
| $-10\sqrt{2}$ | −33.79 | −42.33 | −10 | 2.76 | −2.92 | −36.54 | −39.44 | −33.78 | −42.36 |
| $-5\sqrt{2}$ | −16.14 | −20.65 | −5 | 1.38 | −1.45 | −17.54 | −19.16 | −16.16 | −20.61 |
| $5\sqrt{2}$ | −18.94 | −17.79 | 5 | −1.39 | 1.43 | −17.54 | −19.16 | −18.93 | −17.73 |
| $10\sqrt{2}$ | −39.38 | −36.81 | 10 | −2.80 | 2.83 | −36.54 | −39.44 | −39.34 | −36.61 |

In the following, we present the results related to the variations in the ground state energy levels of the electron and hole upon changing the intensity and orientation of the laser and static EFs. We also studied the variation in oscillator strength of the interband transition and the corresponding radiative recombination time. Regarding the latter, a comparative reference to results reported in other articles for nanostructures of different shapes and sizes, found in various physical conditions, can be useful [42-44].

The results of the analysis in the case of transverse laser dressing of the DQR, with or without a transverse EF, are shown in Figs. 7-9. Figure 7 shows the results obtained as a function of the electronic laser parameter at some selected values of the EF in the situation where both applied fields are oriented perpendicular to the symmetry axis of the DQR (see the medallion in Fig. 7a and the EF notations). We mention here that the values of the ground state energies of the electron and the hole obtained in the absence of any external field are in very good agreement with those reported in Ref. [17]. From Fig. 7a, it is observed that the general tendency of the electron energy is to increase with the laser parameter, which is a well-known effect for the laser dressing of nanostructures [34]. If the applied EF is parallel to the laser polarization direction, the increase in energy is faster for $\alpha_e < 10$ nm. For $10$ nm $< \alpha_e < 16$ nm, the growth rate was approximately the same regardless of the orientation of the EF. If $16$ nm $< \alpha_e < 20$ nm, a second faster growth stage was recorded for the $x$-orientation of the EF. An EF $F_x$ parallel to the LF favors wider variations in electron energy. The same observation is valid for hole, whose energy variations are much smaller and almost insignificant (Fig. 7b). It must be explained here, which is also useful for the following figures, why the energy of the confined particle can become negative in EF. The EF term of the Hamiltonian, that is $-q_{e,h}\boldsymbol{r} \cdot \boldsymbol{u}_f F$, is a positional energy, so it is only relatively determined (its value depends on the choice of the origin of the coordinate system used, which in this work is the center of the DQR basis). Therefore, negative values of the eigenenergies indicate energy levels below the arbitrary reference level chosen as zero energy for each particle. Figure 7c using a logarithmic scale shows that, unless EF is zero, the interband oscillator strength is very low, with small variations that bear the imprint of electron energy variations. A transversely polarized LF cannot practically be used as an oscillator strength control tool if an EF perpendicular to the symmetry axis is simultaneously present. For $F = 0$ the shape of the oscillator strength variation curve bears the imprint of the overlapping integral, as shown in Fig. 6a. Figure 7d also shows the recombination time on a logarithmic scale and demonstrates that the radiative decay occurs efficiently only in the absence of the transverse EF or, to a lesser extent, at relatively low EF values. The LF can nonlinearly control the radiative recombination time in the relevant case $F = 0$, but in a rather small range, between about 1.4 ns and 2 ns.

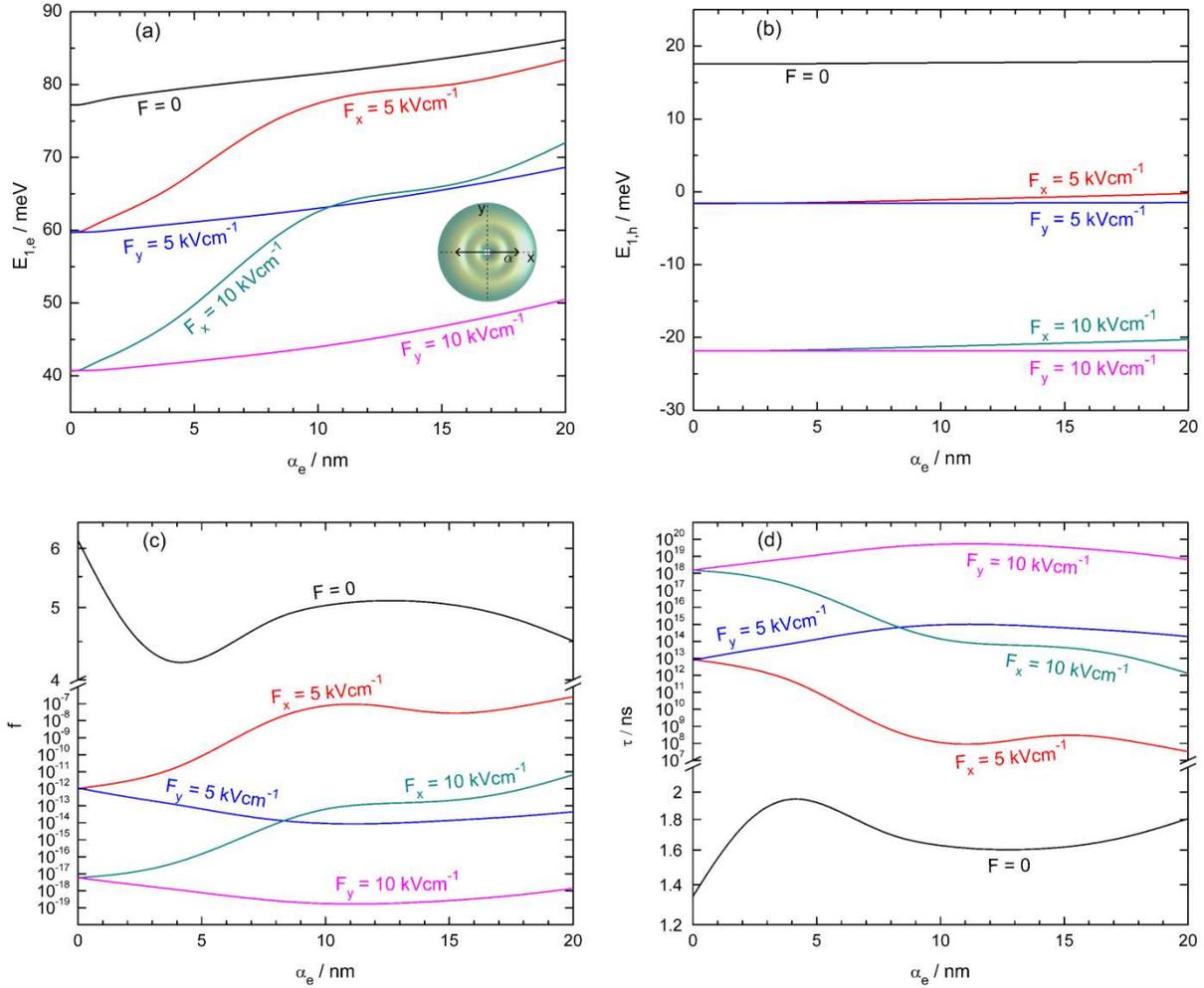

**Figure 7.** Influence of the transversely polarized LF on the electron energy (a), the hole energy (b), the oscillator strength (c) and the recombination time (d), without and with an EF applied parallel or perpendicular to the laser polarization direction.

Figure 8 shows the same results, but as a function of the transverse EF, for two relative orientations with respect to the transversely polarized LF ($\varphi = 0°$ and $\varphi = 90°$), at three values of the laser parameter ($\alpha_e = 0, 10, 20$ nm). Because of the $z$-axis symmetry of the structure and the preservation of the symmetry with respect to the $x$-direction of the laser dressing, it is not necessary to consider the negative values of EF in this case. From Fig. 8a it can be observed that in the absence of the LF or with an LF polarized perpendicular to the EF, the Stark effect of the electron is quasi-linear. The slopes of the graphs were independent of the laser parameter in these cases. In the parallel orientation of the two fields, the Stark effect was quasi-quadratic for $F < 6$ kV/cm and linear for 6 kV/cm $< F < 10$ kV/cm, with a smaller slope at higher laser parameters. In the case of the hole (Fig. 8b), deviation from the linearity of the Stark effect is observed only at small values of EF ($F < 2$ kV/cm). The hole energy is significantly less influenced by the laser parameter and laser polarization direction. Figure 8c shows that the oscillator strength decreases with the applied EF but at very different orders of magnitude, depending on the relative orientation of the two fields. In the perpendicular orientation, the oscillator strength drops off quickly, even at relatively low EF values. This is in agreement with the effective separation of electron from hole in the two deep and diametrically opposed laser-induced confinement zones, as shown in Fig. 2h. All curves in Fig. 8c tend to linearize for strong EFs, which generally indicates an exponential decay of the oscillator strength with static EF strength. Fig. 8d shows that radiative recombination is significant only at EF values below 1 kV/cm.

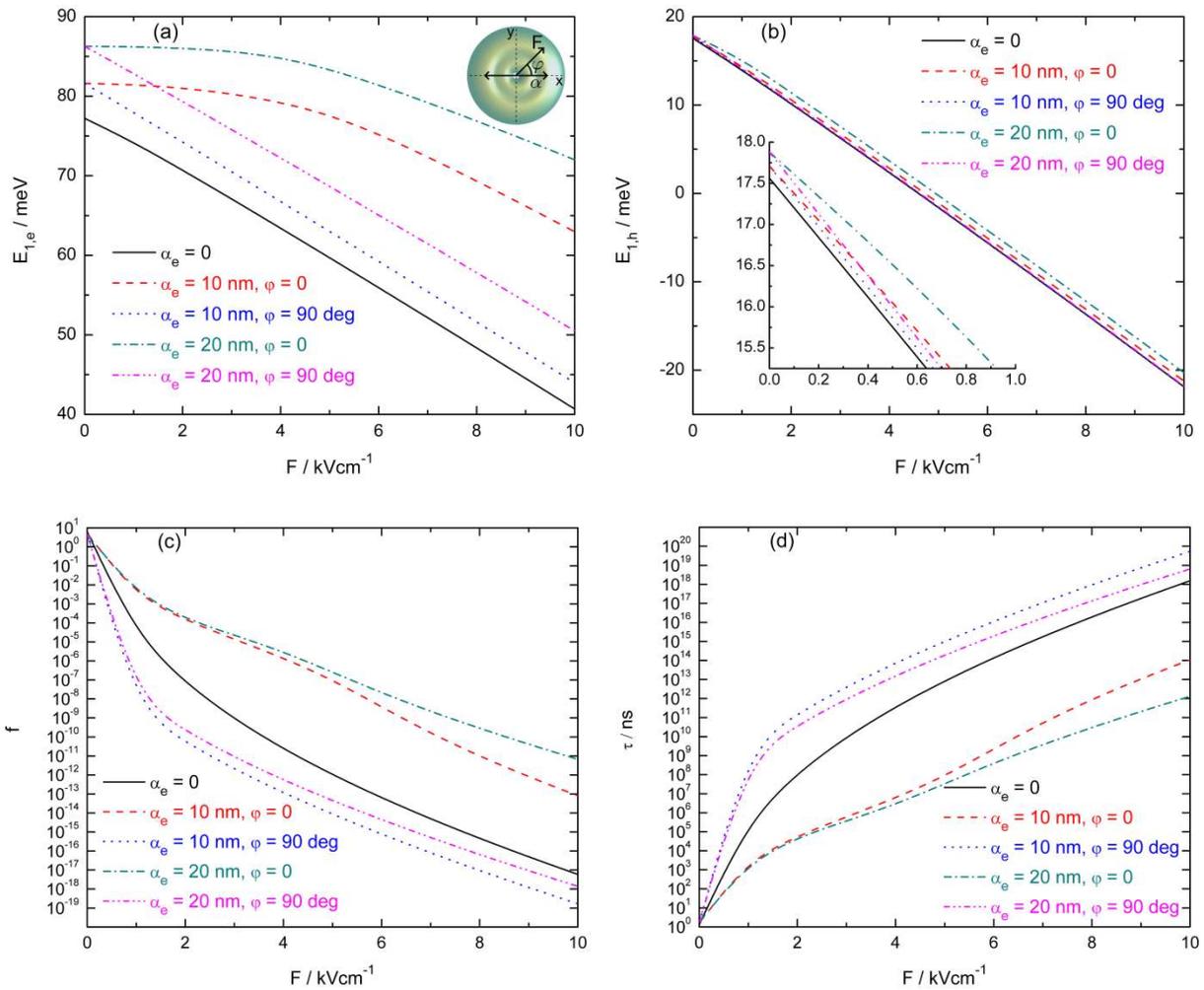

**Figure 8.** The effect of the EF transverse to the $z$ symmetry axis on the electron energy (a), the hole energy (b), the oscillator strength (c) and the recombination time (d), without or with a LF transverse on $z$, parallel or perpendicular to the direction of the EF.

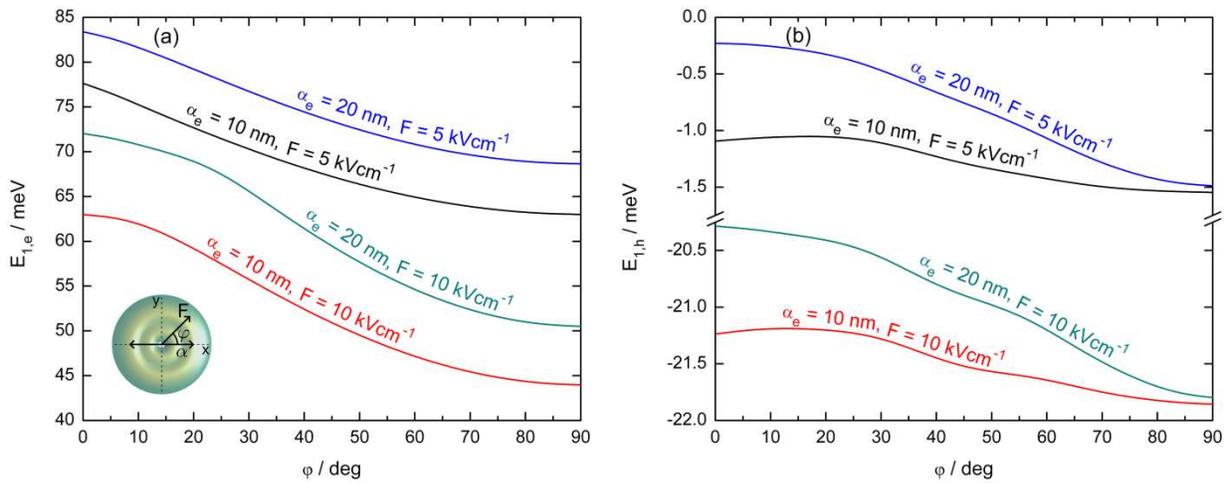

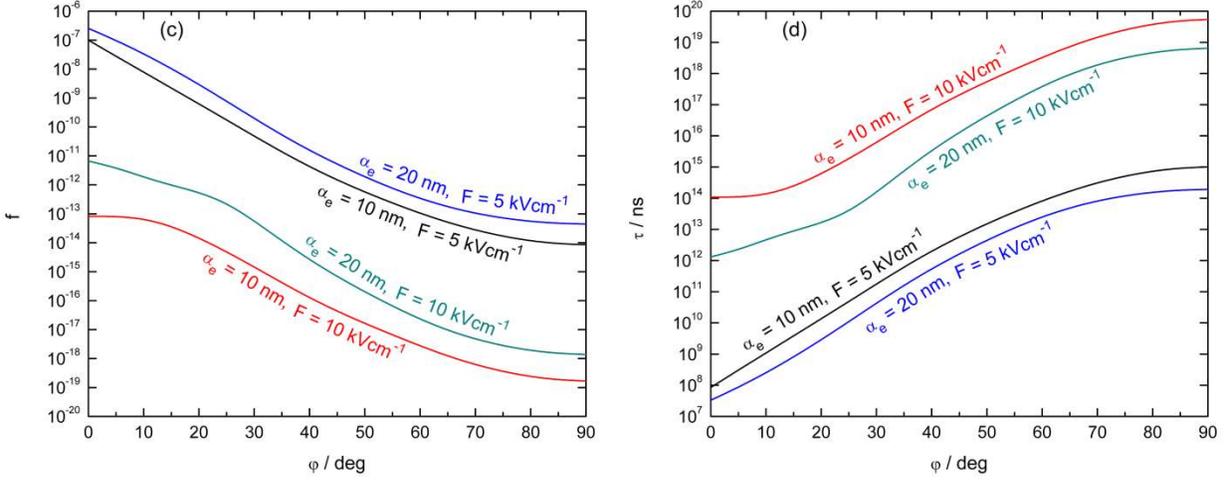

**Figure 9.** Electron energy (a), hole energy (b), oscillator strength (c), and recombination time (d), as functions of the azimuthal angle made by the directions of the LF and the static EF, both of which are transverse to the DQR symmetry axis.

Figure 9 shows the same z-transversal configuration of the fields, but with a variable angle between their directions. Small angles favor slightly higher interband transition energies for both electron and hole, higher oscillator strength values, and lower recombination times.

The axial orientation of one or both fields will result in a very different behavior of the energy levels, with wider variations, as shown in Figs. 10-11. Figure 10 shows the axial laser dressing of the DQR, with or without an axial EF. The increase in electron energy with the laser parameter is strong, as can be seen in Fig. 10a, especially for $0 < \alpha_e < 5$ nm. In this interval, the electron energy increases by approximately 100 meV and is slightly sensitive to the presence and orientation of axial EF. For $\alpha_e < 2$ nm the increase is quasi-parabolic, and then linearizing up to $\alpha_e \cong 4$ nm. In the same range of laser parameters, for the transversal dressing shown in Fig. 7, only a 10 meV increase in electron energy was observed. Figure 10a indicates that the energy increase is more moderate for $\alpha_e > 5$ nm, and therefore the control of the electron energy by intense LF is less effective. For the hole, the energy increase is quasi-parabolic over the entire range considered for the laser parameter, as shown in Fig. 10b. Because the quantum structure is relatively narrow in the axial direction, the laser dressing in this direction changes the hole energy by more than 30 meV. In Fig. 10c, the variations in oscillator strength and recombination time with the laser parameter are plotted together. Although the axial laser dressing can significantly alter the z-axis co-localization of the two particles, by shifting the electron WF to the outside of the quantum structure (see Fig. 5c), the radiative recombination process remains efficient, with moderate increases in the recombination time, with a factor between 50 and 120, depending on the value and orientation of the axial EF.

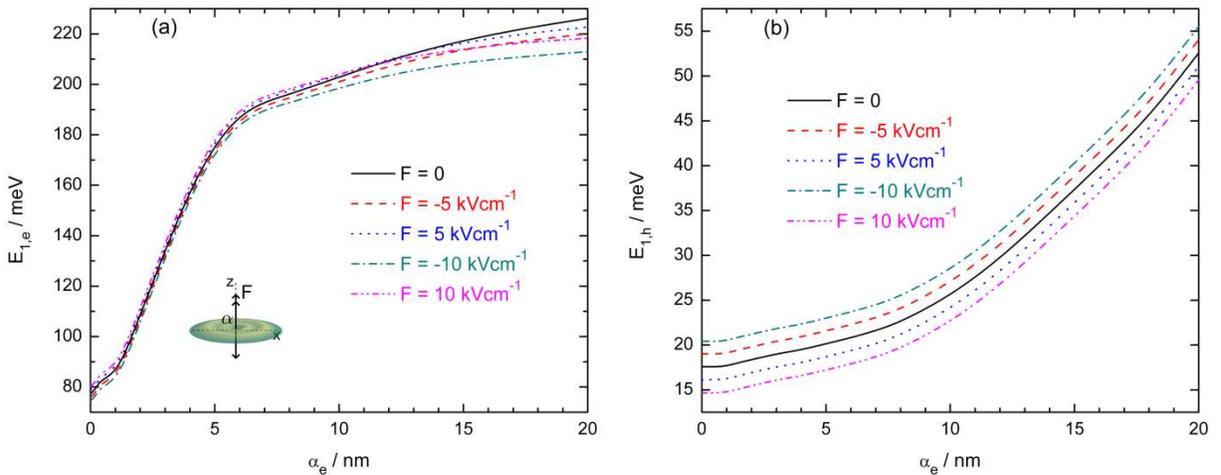

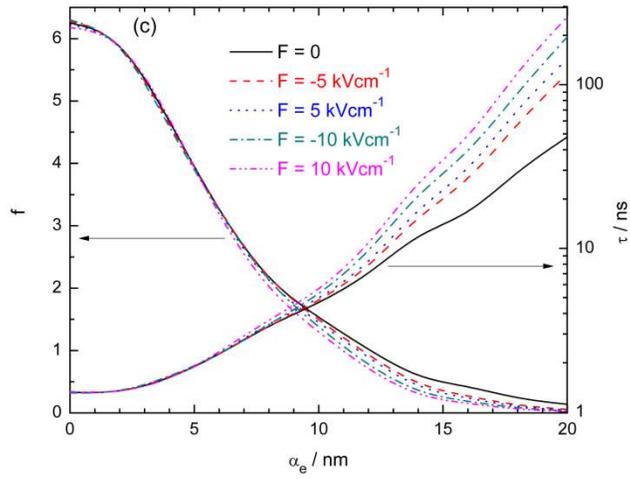

**Figure 10.** Axial laser dressing of the DQR, for several values and orientations of the axial EF: a) variation of the electron energy; b) hole energy variation; c) oscillator strength and recombination time depending on the laser parameter of the electron.

Figure 11 also shows the case of axial fields, but the graphical representations are based on the strength of the applied EF. Here, it is relevant to consider negative values of EF because the quantum structure has no geometric symmetry with respect to any plane transverse to the $z$-axis.

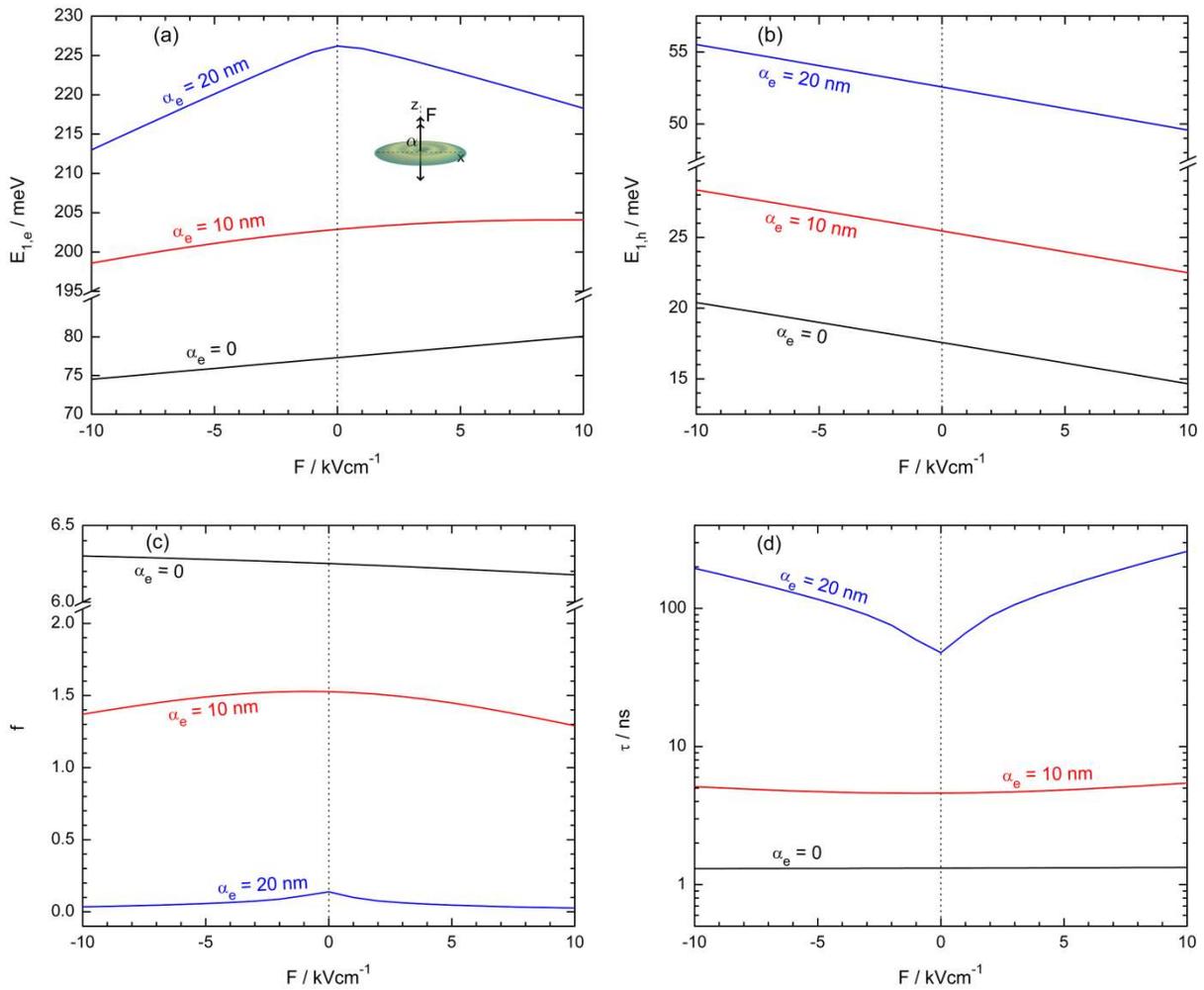

**Figure 11.** Electron energy (a), hole energy (b), oscillator strength (c) and recombination time (d) as a function of the axial EF, at three values of the laser parameter.

From Fig. 11, it is observed that the axial EF has a weak influence on the oscillator strength and recombination time, especially at small values of the laser parameter.

Figure 12 shows the results for the variation of charge carrier energies, oscillator strength, and recombination time as a function of the polar angle $\gamma$ of the laser polarization direction in the absence of the static EF. These results are interesting from the perspective of using the rotation of the polarization direction of constant-intensity laser radiation as a way to control the interband radiative properties. The general trend of the energy levels is to decrease with increasing $\gamma$, an expected behavior given that laser dressing has a greater effect along the directions where the structure has a smaller effective size. In Fig. 12a, one can compare the variations with $\gamma$ in the electron and hole energies at two values of the laser parameter. It is observed that the electron energy decreases relatively slowly up to large values of the angle and very rapidly as we approach the limit case of transverse dressing. In the case of the hole, the inflection of the curves occurs at smaller angles, and the decrease in energy is less extensive. Figure 12b shows, on the same logarithmic scale, the somewhat complicated but small-scale variation of the oscillator strength and recombination time upon the rotation of the laser polarization direction. The mixed monotonicity of the curves is most likely related to the particular geometric profile of DQR.

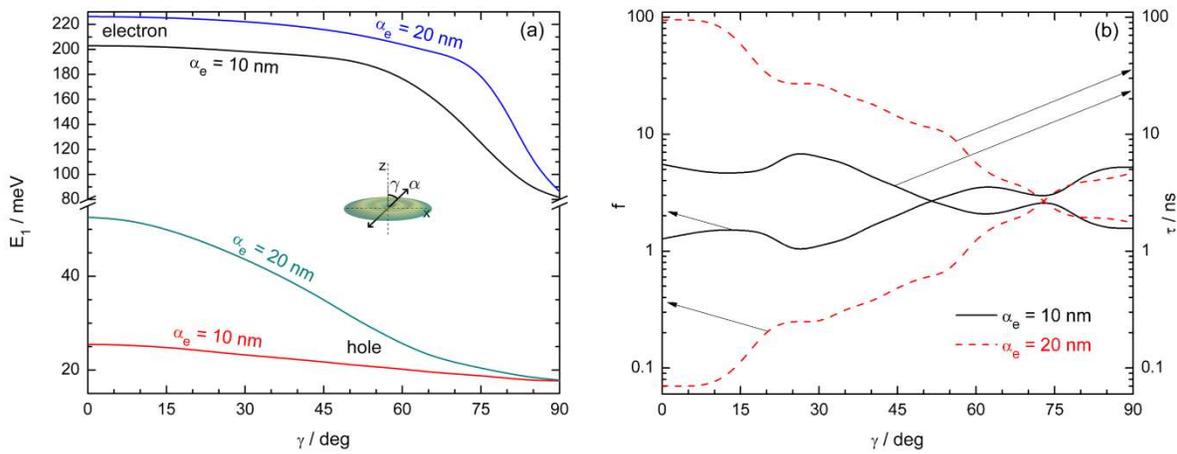

**Figure 12.** Laser dressing with variable polarization direction between the axial case ($\gamma = 0°$) and the transverse case ($\gamma = 90°$), at two values of the laser parameter: a) variation of the energy levels of the electron and the hole; b) variation of oscillator strength and recombination time.

If only an EF acts on the DQR, the behavior of the confined particles is very sensitive to the orientation of the field with respect to the axis of symmetry, that is, there is a strongly anisotropic field-induced polarizability. The strong influence of nanostructure geometry on electric polarizability was also reported in other recent works on the Stark effect [45]. As discussed, the energy levels and especially the recombination time are very different in the two main orientations: axial and transverse. Figure 13 shows the effects of variable polar orientation $\theta$ of EF for two values of its strength. The angle was varied between 0° and 180°, because the nanostructure had no transverse plane of symmetry. It is observed, particularly in Fig. 13a, that the curves are not exactly symmetrical with respect to the dotted line $\theta = 0°$. In this figure, the energy increase, both for the electron and hole, can be noted when the orientation of EF approaches the axis of symmetry in both directions of the axis. In addition, increasing the EF strength has a much greater effect on the variation in the energy levels if the field orientation is close to the transverse direction. In Fig. 13b, are plotted on different logarithmic scales, the oscillator strength of the interband transition, and the corresponding recombination time. The more intense the EF, the sharper the increase of the recombination time from the case of parallel orientation ($\theta = 0°$) or antiparallel orientation ($\theta = 180°$) to the case of transverse orientation ($\theta = 90°$). These results are explained by the significantly sharper spatial separation of the electron and hole WFs when EF is oriented close to the transverse plane of the DQR.

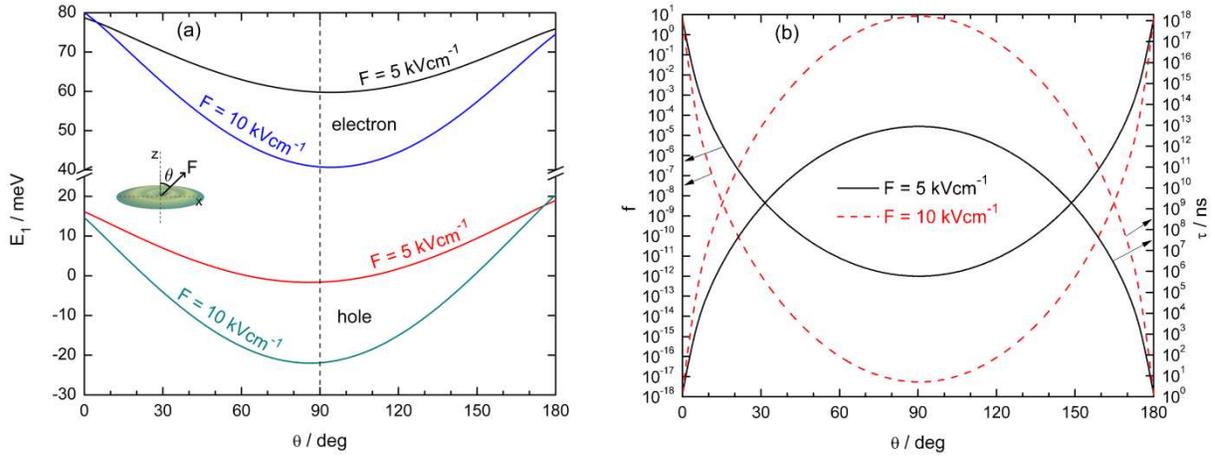

**Figure 13.** Dependence of the Stark effect on the polar angle $\theta$ of the applied EF, at two values of the field: a) variations of the energy levels of the electron and the hole; b) variations of the oscillator strength and electron-hole recombination time.

Because the tilted laser dressing of the energy levels cannot be interpreted as a superposition of the axial and transverse dressings (see Table II), it is important to study the behavior of the charge carriers with an increase in the laser parameter at different orientations $\gamma$ of laser polarization. The results are shown in Fig. 14. The electron energy increased monotonically with the laser parameter, regardless of the field orientation (Fig. 14a). What changes with $\gamma$, however, is the magnitude of the growth, which is much larger near the axial orientation, as well as the range of $\alpha_e$ values in which the growth is rapid. It is observed that the small-angle curves in Fig. 14a show some inflections at small values of the laser parameter: their shape is rather convex for $\alpha_e < 3$ nm. This also explains the convex shape of the hole energy growth curves in Fig. 14b, bearing in mind that the laser dressing parameter of the hole is much smaller than that of the electron at the same laser intensity. Figures 14c and 14d show the variations with $\alpha_e$ of the oscillator strength and recombination time, respectively. The oscillating behavior of the curves in Fig. 14c is less intuitive and is related to the particular spatial configuration of the DQR confinement region. In Fig. 14d, it is observed that the recombination time is significantly more affected by the increase in the laser parameter at small values of the angle $\gamma$. Although the monotonicity of the curves was mixed, the overall trend of the recombination time was to increase with $\alpha_e$.

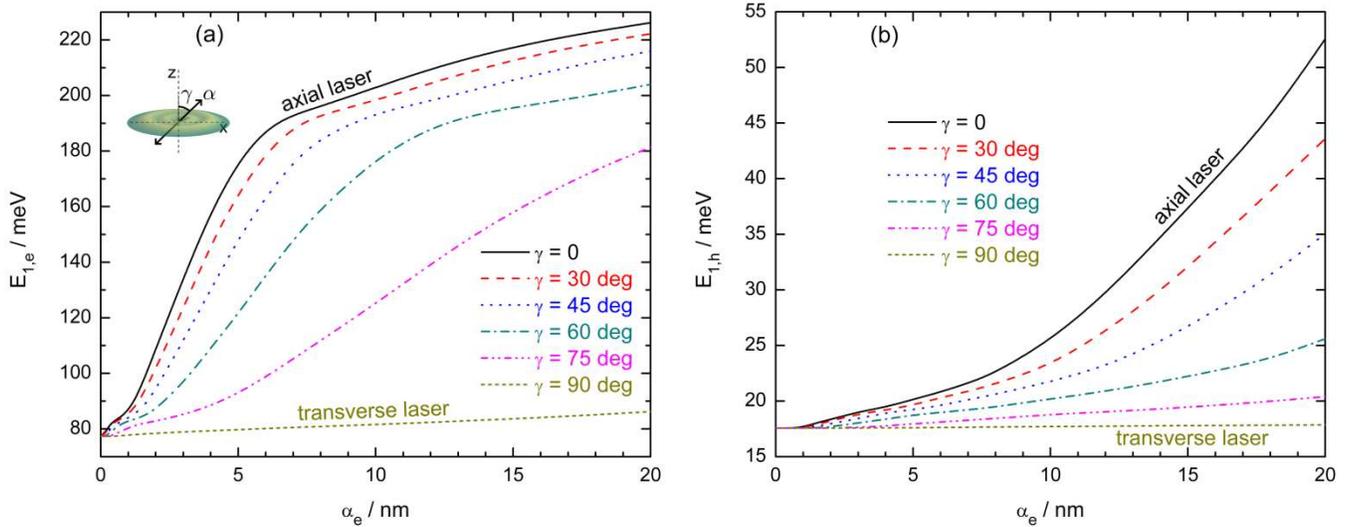

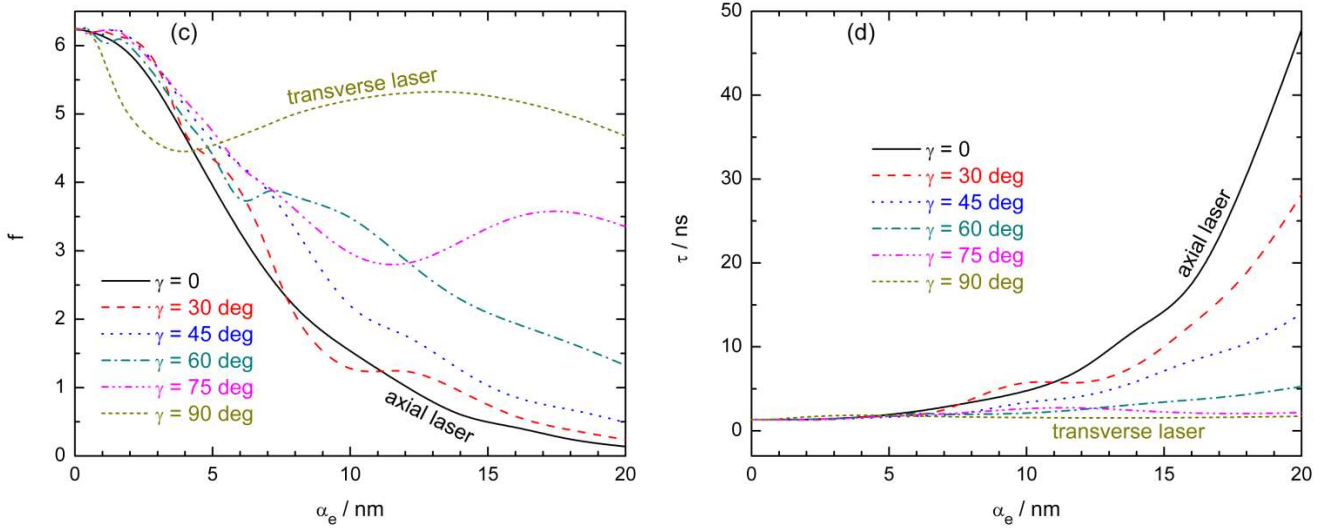

**Figure 14.** Laser dressing in several directions tilted with respect to the axis of symmetry of the DQR. a) electron energy; b) hole energy; c) oscillator strength; d) recombination time, as functions of the laser parameter.

It is useful to understand the extent to which the LF and the static EF can be used as instruments for tuning the vacuum wavelength $\lambda = ch/E_t$ of the radiation absorbed/emitted by the DQR through the interband transition. In order to have applicative meaning, this information about the variation of the wavelength must be related to the corresponding variation of the oscillator strength, in the same range of laser parameter or electric field.

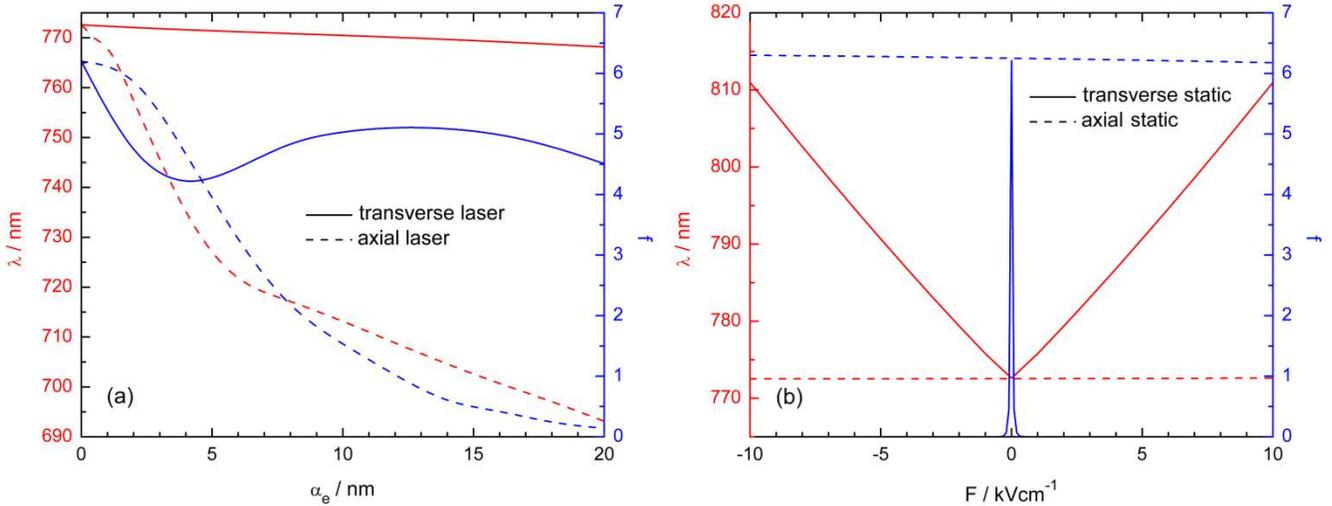

**Figure 15.** Interband transition wavelength (red lines) and oscillator strength (blue lines) as functions of the laser parameter of the electron (a) and static electric field (b). The two main orientations of the fields were considered.

Figure 15a shows the change in wavelength and oscillator strength with the electronic laser parameter, for the two main LF polarizations. It is observed that the transversely polarized laser, although it maintains relatively high values of the interband oscillator strength in the entire range of $\alpha_e$ values, can only ensure a minor decrease in the emitted wavelength. Regarding the axial polarization, the tuning of the wavelength is much wider, from approximately 773 nm to 693 nm, thus with a blueshift of about 80 nm. Although in this case the oscillator strength also decreases from approximately 6.20 to 0.14, its values are still significant for the tuning mechanism to be of practical interest. Figure 15b presents the change in wavelength and oscillator strength with the static EF, for its two main orientations. It is observed that the transverse EF, although it produces a significant Stark redshift from approximately 773 nm to 811 nm, it cannot be a practical way of tuning due to the extremely rapid decrease to zero of the oscillator strength. In the case of the axial EF, the oscillator strength theoretically keeps almost constant its high value. However, in this case there is practical no tuning of the wavelength, because the confined Stark effect is very weak at all considered values of the field. Figure 15 therefore demonstrates three major aspects of practical importance: i) the axially-polarized non-resonant LF is a good instrument for tuning the interband wavelength; ii) the transversely-oriented static EF can be very effective to control the oscillator strength and implicitly the interband recombination time; iii) there is a strongly anisotropic behavior of the quantum structure with respect to external fields.

Potential applications could exploit the DQR anisotropy in the design of polarizers, selective sensors and radiation detectors, optical switches and new optical memory cells (given the drastic switching of the recombination time in transverse electric fields).

## 4. Conclusions

We theoretically studied a DQR model under the separate or simultaneous action of a static EF and an intense non-resonant LF. To the best of our knowledge, this is the first 3-D model that considers the exact laser dressing along different directions of a realistic quantum structure. We analyzed how the interband transitions are modifiable through the interaction of the DQR with the two types of fields. Both interactions were investigated in detail at variable intensities of the applied fields and at various orientations of the fields with respect to the axis of symmetry of DQR. The difference in behavior between the effects of the axial and transverse fields was observed. We have shown that the Stark shift of the energy levels when the EF is oriented in a certain direction can be deduced from the individual shifts when the field components are applied in the axial and transverse directions. However, this is not true for laser dressing in a tilted direction. Our results demonstrate that the interband recombination time can be significantly increased by a transverse EF, and that the electron and hole energy levels can be significantly altered by an axially polarized LF and as a consequence this type of field is effective for tuning the wavelength of the interband transition. The electric polarizability of the electron/hole pair was found to be highly anisotropic. In addition, at a given relatively large value of the laser parameter, the rotation of the laser polarization direction produces a sensitive variation in the recombination time. These results can be useful for practical applications in which interband photoluminescence controlled by external fields is required.